\documentclass[prb,twocolumn,showpacs,preprintnumbers,amsmath,amssymb,groupeaddress,superscriptaddress,floatfix]{revtex4}

\usepackage{graphicx}
\usepackage{dcolumn}
\usepackage{bm}
\usepackage{amsmath}
\usepackage{pstricks}

\newcommand{\VEV}[1]{\left\langle {#1}\right\rangle} 
\newcommand{\Bra}[1]{\left\vert {#1}\right\rangle}
\newcommand{\abs}[1]{\vert {#1}\vert}

\def\R{\mbox{\boldmath $r$}}
\def\SS{\mbox{\boldmath $S$}}
\def\H{\mbox{\boldmath $H$}}
\def\B{\mbox{\boldmath $B$}}

\def\s{\hbox{\rm s}}
\def\tesla{\hbox{\rm T}}

\def\muB{\hbox{$\mu_B$}}
\def\K{\hbox{\rm K}}
\def\Cot{Co$^{3+}$}
\def\Cod{Co$^{2+}$}
\def\Coq{Co$^{4+}$}
\def\Nuc{$^{59}$Co}

\def\CaCO{Ca$_3$Co$_2$O$_6$}

\newcommand{\commentout}[1]{}

\def\addrParma{Dipartimento di Fisica e Unit\`a CNISM, Universit\`a degli Studi di Parma, Viale G.\ Usberti 7A, I-43100 Parma, Italy}
\def\addrWarwick{Department of Physics, University of Warwick, Coventry, CV4 7AL, United Kingdom}
\def\addrCaen{Laboratoire CRISMAT, UMR 6508, Boulevard du Mar\'echal Juin, 14050 Caen Cedex, France}
\def\addrDresden{Max Planck Institute for Chemical Physics of Solids, N\"othnitzerstr. 40, 01187 Dresden, Germany}
\def\addrESRF{European Synchrotron Radiation Facility, BP 220, 38043 Grenoble Cedex 9, France}
\def\addrMilano{Dipartimento di Fisica, Politecnico di Milano,
Piazza Leonardo da Vinci 32, I-20133 Milano, Italy}

\begin{document}

\title{Magnetic order, metamagnetic transitions, and low-temperature spin 
freezing in 
Ca$_3$Co$_2$O$_6$: an NMR study.}

\author{G. Allodi}
\email{allodi@fis.unipr.it}
\affiliation{\addrParma}
\author{R. De Renzi}
\affiliation{\addrParma}
\author{S. Agrestini}
\affiliation{\addrCaen} \affiliation{{ \addrDresden}}
\author{C. Mazzoli}
\affiliation{\addrESRF} \affiliation{{ \addrMilano}}
\author{M.~R.~Lees} \affiliation{{ \addrWarwick}}
\date{\today}

\begin{abstract}

We report on a $^{59}$Co NMR investigation of the trigonal cobaltate \CaCO\
carried out on a single crystal, providing precise determinations of the 
electric field gradient and   
chemical shift tensors, and of the internal magnetic fields at the non-magnetic Co\,I 
sites, unavailable from former studies on powders. The magnetic field-induced 
ferri- and 
ferromagnetic phases at intermediate temperature (e.g.\ 10\,K) are identified
by distinct internal fields, well accounted for by purely dipolar interactions.
The vanishing transferred hyperfine field at the Co\,I site indicates that 
the \Cot\,(I) orbitals 
do not participate in the intra-chain superexchange, in disagreement with 
 a previous theoretical model.

The strong Ising character of the system is confirmed experimentally by 
the field dependence of the resonance lines, indicating that local 
moments are saturated even at the phase boundaries. 
In the vicinity of the critical fields, nuclear 
spin-spin relaxations
detect the 
spin reversal dynamics of large magnetic assemblies, either Ising chain 
fragments or finite size domains, which drive 
the metamagnetic transitions. Such collective excitations 
 exhibit a glassy behavior, slowing down to subacoustic  
frequencies and freezing at low temperature.
The relevance of such slow fluctuation modes for 
the low-temperature multi-step behavior reported in the magnetization is 
discussed.
  
\end{abstract}

\pacs{
75.30.Kz, 	
75.30.Et, 	
75.60.Jk, 	
76.60.-k. 
}

\maketitle
\section{Introduction}

\begin{figure}
\includegraphics[width=\columnwidth]{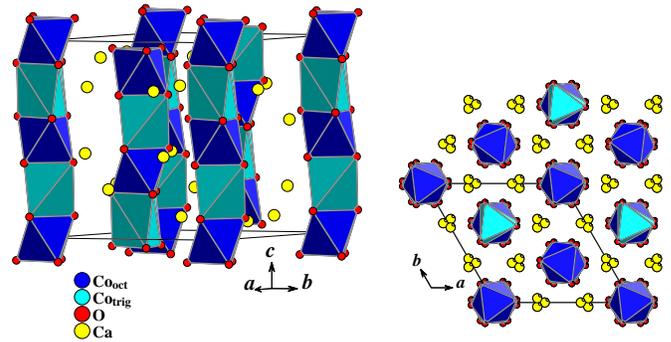}
\caption{\label{fig:struct}
 {(Color online) 
Lattice structure of \CaCO\ in the hexagonal setting. 
Left: perspective view of the spin 
chains formed by the CoO$_6$ polyhedra along the c-axis; right: cross-section 
in the $ab$ plane, evidencing the arrangement of chains in the  
triangular lattice.
The dark 
and light 
polyhedra represent the CoO$_6$ octahedra and CoO$_6$
trigonal prisms, respectively.
}}
\end{figure}

The nature of magnetic order in the geometrically frustrated 
trigonal cobaltate \CaCO, an Ising-type magnet exhibiting low-dimensional 
behavior, 
has been a puzzle for over a decade. 
The crystal structure of \CaCO\ (Fig.\ \ref{fig:struct}) may be viewed as the 
result of the 
alternate stacking along the $c$ axis of
face-sharing CoO$_6$ octahedra (Co\,I) and CoO$_6$ trigonal prisms (Co\,II)  
with a rather short Co\,I - Co\,II distance $c/4=2.59$\,\AA, building up 
{\it chains} arranged over the $ab$ plane in a triangular lattice, with a 
much larger spacing (inter-chain distance $a/\sqrt{3}=5.24$\,\AA). 
\cite{jsolstatechem, Maignan00}
The \Cot\ oxidation state at both sites is now established, 
\cite{powder_nmr, burnus} contrary to earlier predictions for a \Cod-\Coq\ 
charge disproportionation. \cite{vidya}

Magnetism resides on the Co\,II site, where a trigonal crystal field yields a high-spin 
ground state with a large unquenched orbital component ($\VEV{L_z}\approx 1.7$,
 $\VEV{2S_z + L_z}=5.3$), while 
Co\,I is in a low-spin  state
(i.e.\ non-magnetic). \cite{burnus}
Spin-orbit coupling gives rise to a single-ion anisotropy term $-DS_z^2$ of 
order 70~meV (corresponding to an anisotropy field $B_{anis}\approx 200$\,T 
along $c$), whence the strong Ising-type character of magnetism in this 
system. \cite{Maignan00, huawu}  
Exchange coupling  between the \Cot\,II ions is ferromagnetic 
(FM) and relatively strong along the chains, whereas it is antiferromagnetic (AF) 
and much weaker between chains. 
While the dominant intra-chain exchange interaction promotes quasi 
one-dimensional Ising-type ferromagnetism, the actual magnetic structure is 
the result of the competing residual inter-chain AF 
coupling, giving 
rise to a complex  magnetic phenomenology
as a function of temperature  and magnetic field. 


Susceptibility measurements actually demonstrate the onset of magnetic order
 at $T < T_c\approx 25$~K. \cite{earlyneutrons_FI, Maignan00}
The nature of magnetic ordering 
however, has been the subject of a long-lasting controversy. 
Denoting spin-up, spin-down, or magnetically disordered Ising chains 
as $\uparrow$\,,\,$\downarrow$\,,\,0, respectively,
 theoretical models indicated either a ferrimagnetic (FI) 
$[\uparrow\,\uparrow\,\downarrow]$ or a 
 partially disordered antiferromagnetic (PDA) structure 
$[\uparrow\,\downarrow\mskip -2mu 0]$ 
as the stable zero-field arrangements of the three inequivalent chains on the 
triangular lattice 
in the presence of geometrically frustrated AF 
inter-chain coupling.\cite{mekata,wada} 
Early neutron diffraction experiments seemingly supported either a 
PDA \cite{earlyneutrons_pda} or a FI \cite{earlyneutrons_FI} order, 
although poor quantitative agreement of both structures with magnetic peak 
intensities  was later pointed out. \cite{earlyneutrons_neither}
Recent 
x-ray \cite{agrestini_RXS} and neutron scattering \cite{agrestini_prl} data 
on single crystals have revealed the incommensurate character of the magnetic 
order in this compound.
They show that
the actual structure,
referred to as {\em modulated PDA} (MPDA), rather 
consists of 
long-wavelength longitudinal spin-density modulations along each chain, phase-shifted by 
$\pm2\pi/3$ with respect to the neighboring chains. \cite{agrestini_prl}
Nevertheless, the MPDA phase is apparently metastable 
 below 14\,K,  
as indicated by the decrease in the intensity of the MPDA magnetic peaks. 
\cite{earlyneutrons_neither,agrestini_RXS,agrestini_prl}

The nature of the magnetic order in an externally applied magnetic field is 
also far from clear. 
Increasing magnetic fields applied along $c$ tend to stabilize the FI phase 
and eventually FM inter-chain order. At 
$5\,\K< T < T_c$, 
the initial
 saturation of the magnetization to a magnetic moment 
$M_{FI}\approx 1.7\,\muB$ per formula unit at $\mu_0H > B_{c1}\!\approx\! 0.1\,\tesla$, 
followed by a step to 
$M_{FM}\approx 3M_{FI}$ at $\mu_0H > B_{c2}\! \approx\! 3.5\,\tesla$, 
actually points to 
perfect FI and FM phases below and above $H_{c2}$, respectively. 
However, the $M(H)$ curves at lower temperatures 
display a puzzling multi-step behavior reminiscent of the quantum tunneling of
magnetization detected in molecular magnets. \cite{jmatchem,Hardy04}
A number of complex magnetic superstructures, 
distinct from both the FI and FM
spin-chain arrangements, have been proposed in order to explain the fractional
magnetization values. However, such superstructures cannot account 
for all the plateaus observed in $M(H)$. \cite{superstruct}

In this paper, we address the issue of the field-dependent magnetic order of
\CaCO\ by means of \Nuc\ nuclear magnetic resonance (NMR). 
Previous \Nuc\ NMR research on \CaCO\ polycrystalline samples, 
though successful in demonstrating the presence of non-magnetic \Cot, 
could only provide very limited information on the 
magnetic structure, due to the angular averaging of the complex multi-line 
quadrupolar spectra into broad powder patterns. \cite{powder_nmr} 
The FM and FI surroundings, on the contrary, can be clearly resolved in our 
NMR investigation of a high-quality single crystal, while
the two field-induced transitions, 
namely MPDA-FI and 
FI-FM, were precisely identified  at 10\,K through     
step-like jumps in the internal field and peculiar features   
of the spin-spin relaxations, indicative of glassy spin dynamics. 
Our study also yields a 
detailed characterization of the 
chemical shift and the electric 
field gradient (EFG) tensors at the non-magnetic  Co\,I site. 

The paper is organized as follows. Sample 
preparation and the experimental methods are briefly described in 
section \ref{sec:experimental}. 
NMR spectra, 
providing 
information on the magnetic structures, and nuclear relaxation data, 
probing the 
dynamical excitations of the spin system, are presented in sections 
\ref{sec:results.spectra}, and \ref{sec:results.relax}, 
respectively.
The calculated dipolar fields at the Co\,I sites in the FI and FM phases are 
compared 
with experimental values in section \ref{sec:dipfield}.
The results are discussed in section \ref{sec:discussion}. 
Two appendices 
review in some detail
the complex dependence of the \Nuc\ resonances on magnetic and quadrupolar 
interactions, and the longitudinal relaxation function of a $7/2$ nuclear spin. 

\section{Experiment}
\label{sec:experimental}

 
The investigated sample was a single crystal of size 
$1.2 \times 1.2 \times 5.8$ mm$^3$ (the long side parallel to 
the $c$ axis) taken from a batch of several needle-shaped crystals obtained by 
the flux method.
The crystals were grown using the following procedure: a mixture of \CaCO\ 
powder and KCO$_3$ (the flux), in a weight ratio 1/7, was heated up to 
990$^\circ$\,C in an alumina crucible for one hour and then slowly cooled down 
to room temperature. A sample of the same batch was employed for the resonant 
x-ray scattering study of Ref.\ \onlinecite{agrestini_RXS}.
The high quality of the crystals
was confirmed by preliminary x-ray diffraction, energy dispersive
x-ray, magnetization, and specific heat measurements.

Magnetization measurements were performed after zero field cooling, by using 
a vibrating sample magnetometer (Oxford Instruments) equipped with a 12\,T 
magnet, with a 
field sweep rate of 1\,T/min in order to minimize any 
relaxation effect.
Susceptibility measurements were carried out by means of a superconducting 
quantum interference device magnetometer (SQUID) 
with magnetic fields up to 5\,T and temperatures down 
to 1.8\,K (Quantum Design MPMS). The magnetic measurements were carried out 
for the two geometries 
corresponding to the magnetic field applied parallel or perpendicular to 
the $c$ axis, i.e.\ parallel or perpendicular the direction of the chains
and the spins.
In the case of 
parallel geometry the crystals were aligned using the magnetic field itself 
at 5\,T as explained in Ref.\ \onlinecite{Hardy03}. 
The alignment of the crystals in the perpendicular geometry can be easily 
obtained thanks to their rod-like shape. 

The NMR experiments were performed by means of a 
home-built phase-coherent pulsed spectrometer \cite{hyrespect} and a 
fast-sweeping 
cold-bore cryomagnet (Oxford Instruments EXA) equipped with a variable 
temperature insert as a sample environment.
The crystal was mounted on a sample rotator, providing an angular span of 
$\pm120^\circ$ with respect to the applied field. 
Misalignment of the rotation axis (nominally perpendicular to the field)
was estimated not to exceed 5 degrees. 
In the case of the $c$ axis parallel to the field, however, a 
precise alignment of the sample was achieved, thanks to its large magnetic 
anisotropy, by loosely mounting it in the coil.

NMR spectra were detected by exciting spin echoes 
by means of a 
standard $P-\tau-P$ sequence, with equal rf pulses $P$ of 
duration 3-5 $\mu s$ and intensity suitably regulated for optimum signal, and 
delays $\tau$ kept as short as possible with respect to the dead 
time of the resonant probehead ($\approx $10-25$ \mu$s depending on the 
working frequency).
Recording was carried out 
either by tuning the probehead at discrete 
frequencies in a constant field (frequency sweep mode), or by varying the 
applied field and exciting the resonance at a constant frequency (field sweep 
mode). 
Although 
an indirect method, we preferentially employed the latter whenever possible 
(namely, far from the metamagnetic transitions) as it yields smoother data, 
independent of the frequency response of the spectrometer, in a  
fully automated procedure.

Nuclear spin-spin relaxations were determined by the decay of the signal 
amplitude in 
the same spin echo sequence $P-\tau-P$, as a function of variable delay
$\tau$. Spin-lattice relaxations were measured by the signal recovery 
following a so-called fast saturation of the observed nuclear transition  
(see Appendix \ref{sec:redfield}), 
obtained with an aperiodic train of 10-15 pulses.  


\section{Experimental results}
\label{sec:results}


\subsection{NMR spectra}
\label{sec:results.spectra}

We report separately the \Nuc\ resonance spectra recorded in the 
longitudinal geometry (external field $\B_{ext} \parallel c$ axis),
 a geometry that allowed us 
study the field-induced magnetic order in the system, and in transverse 
applied fields.
Interpretation of the spectra is relatively simple and model-independent in 
the longitudinal case, 
due to the collinearity of all the fields (internal and 
external magnetic fields, and the EFG). 
In the general case, the dependence of the \Nuc\ resonances on magnetic and 
quadrupolar interactions is rather complex, as recalled in some detail in  
Appendix \ref{sec:nuc}.

\vspace{-1em}
\subsubsection{Longitudinal applied fields}
\label{sec:results.spectra.long}
\vspace{-1em}

A \Nuc\ NMR signal could be detected below 15~K in moderate 
longitudinal external magnetic fields $B_{ext} > 0.1$\,T. 
Such field values assign the system to the FI phase. 
For such field values the FI phase is induced in the system.
In contrast, no signal could be detected in the MPDA phase 
at $B_{ext} < 0.1$\,T. 
 
\begin{figure}
\includegraphics[width=\columnwidth]{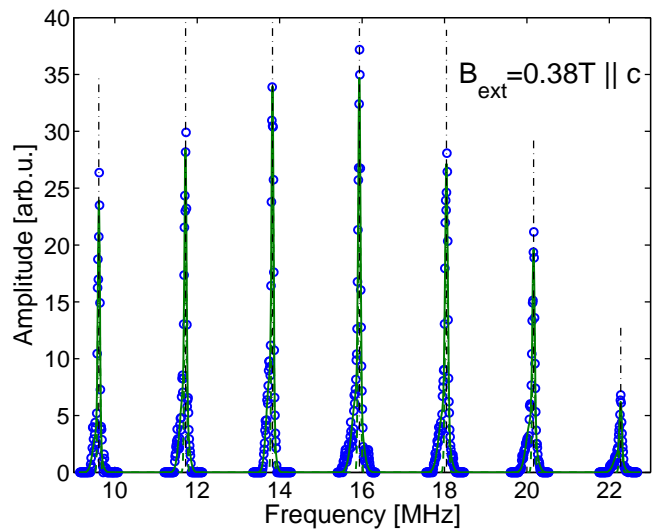}
\caption{(Color online) \label{fig:fsFI}
Frequency-swept \Nuc\ spectrum at 10~K from the majority spin-up chains of the 
FI phase, recorded in a 
field of 0.38~T applied  
along the $c$ axis. Solid lines are bimodal fits of the satellite peaks, 
vertical
dash-dotted lines mark their centers of gravity.}
\end{figure}

A typical frequency-sweep spectrum, recorded at 10~K in 0.38~T, 
is plotted in Fig.\ref{fig:fsFI}. The spectrum consists in  a septet of
quadrupole-split Zeeman  transitions 
$\nu_n$, $n=-3,\dots, n=3$ (Eqn.~\ref{eq:nuQm}), 
with constant frequency spacing $\nu_{n+1} - \nu_n= 2.11(1)$~MHz, in 
agreement with Ref.\ \onlinecite{japan_nmr}. 
In this 
geometry ($\theta=0$), and in the presence of a cylindrical EFG ($\eta = 0$), 
as is actually the case (see Sec.\ \ref{sec:results.spectra.trans}), 
such a line spacing coincides with
the quadrupolar coupling parameter $\nu_Q$ (Eqn.~\ref{eq:ABC}).
The resonance frequency $\nu_0$ of the central line, which is unaffected by 
the quadrupole interaction, corresponds to a field 
$B_{nuc} = 2\pi\nu_0/^{59}\gamma$ at the nucleus which is 
larger than $B_{ext}$  by approximately 1.2~T (here, 
${^{59}\gamma}=2\pi\times 10.10$\,MHz/T \cite{ref_on_gamma}
is the gyromagnetic ratio 
of \Nuc). Such a 
field offset, though sizable, is nevertheless much smaller than the typical 
hyperfine fields detected in magnetic \Cot,  which are of the order of 
 -10~T/$\mu_B$ and 
+60~T/$\mu_B$ for the spin and orbital hyperfine components, respectively. 
\cite{pieper_erco,yoshie}
This confirms that the present \Nuc\ signals are from 
\Cot\ ions in a low-spin state, 
in agreement with previous NMR reports. \cite{powder_nmr,japan_nmr}
We note that the dipolar field at site I generated by two nearest-neighbor 
Co\,II moments of 
5.2~$\mu_B$ aligned parallel to $\B_{ext}$ accounts for both 
the positive sign of the field offset $B_{nuc}-B_{ext}$, and its magnitude to  
within a few percent accuracy. 
This is the first indication that the 
internal field $\B_{int}\approx \B_{nuc}-\B_{ext}$ arises 
from the surrounding 
magnetic ions essentially through the dipolar coupling, as actually 
proven in detail by the internal field calculations of 
Sec.\ \ref{sec:dipfield}.
Given its dominant on-chain dipolar origin, the positive sign 
of $B_{int}$ then indicates that the nearest-neighbor \Cot\ spins 
are aligned along the direction of 
the applied field.
The present positive-offset line multiplet therefore originates from nuclei at 
the Co\,I sites of spin chains aligned parallel to the external  
field, namely, the majority chains of the 
FI structure.

\begin{figure}
\includegraphics[width=\columnwidth]{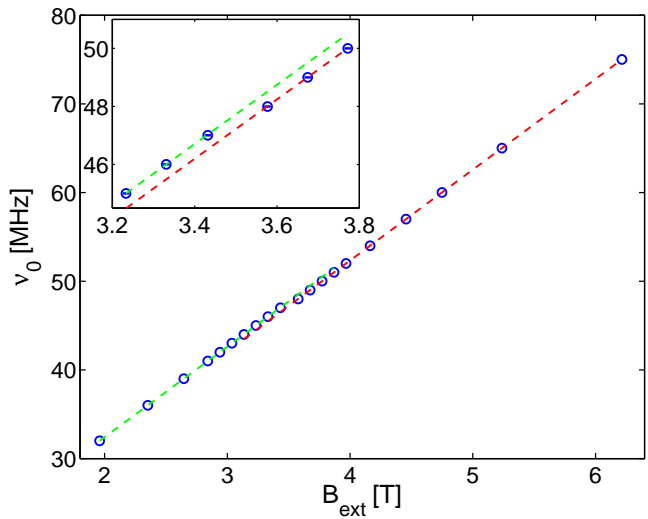}
\caption{(Color online) \label{fig:gamma}
Central resonance frequency $\nu_0$ of the positive-offset spectra from the 
spin-up chains in either the FI ($B_{ext}< 3.5\,\tesla$) or the
FM phase ($B_{ext} > 3.5\,\tesla$), as a function of  $\B_{ext} \parallel c$.
The dashed lines are fits to straight lines for the two phases. The inset is a 
blow-up of the transition region.}
\end{figure}

Qualitatively similar positive-offset spectra could be reproduced
in increasing external fields 
parallel to the $c$ axis ($\theta=0$) at 10~K, up to 
$B_{ext}=3.5\,\tesla$.
The resonance frequency of the central line follows a linear dependence on 
$B_{ext}$ (Fig.~\ref{fig:gamma}),
\begin{equation}  
\label{eq:linear}
\nu_0 = aB_{ext} + b
\end{equation}  
with best-fit parameters 
$a_{\uparrow}=10.21(1)$~MHz/T,  
$b_{\uparrow}=12.02(2)$~MHz (hereafter, $\uparrow$ $\downarrow$ in the subscripts
denote quantities related to spin-up majority and spin-down minority chains of 
the FI phase, respectively).
On the other hand
\begin{equation}  
\label{eq:linearB}
\nu_0 \equiv {^{59}\gamma}B_{nuc} =
{^{59}\gamma}(1+K_c)(B_{ext} + B_{int})
\end{equation}   
 as detailed in Appendix \ref{sec:nuc}. 
If the staggered magnetization 
is saturated all over the field span,
the internal field $\B_{int}$ is independent of $B_{ext}$ and, by symmetry, 
also parallel to $c$.
Working under this hypothesis, whose validity is demonstrated in the 
discussion, a comparison of Eqs.~\ref{eq:linear} and \ref{eq:linearB} 
simply relates the coefficients $a$, $b$ to 
$B_{int}$ and the $c$ axis component of the chemical shift tensor $K_c$:
 $a={^{59}\gamma}(1+K_c)$, $b={^{59}\gamma}(1+K_c)B_{int}$, respectively.
Best-fit values for $a_{\uparrow}$ and $b_{\uparrow}$ then 
yield $B_{int\uparrow}=1.178(2)$~T, and a sizable
$K_c\approx 0.01$.

\begin{figure}
\includegraphics[width=\columnwidth]{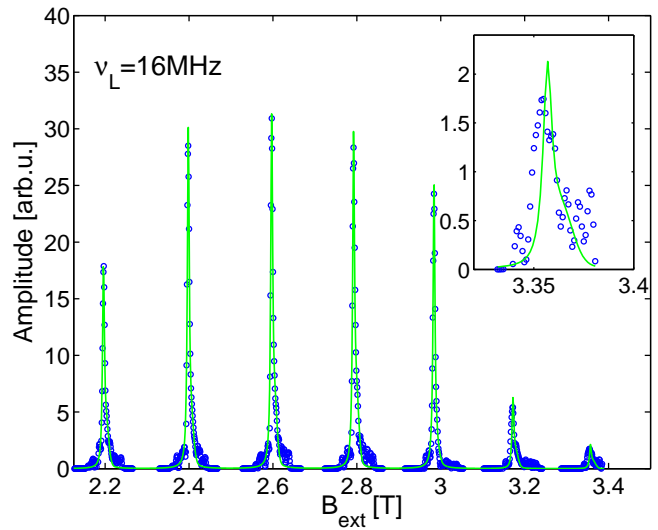}
\caption{(Color online)\label{fig:hsFIrev}
Field-swept spectrum ($\B_{ext}\!\parallel\! c$) at 10~K from the minority 
spin-down 
chains, recorded at a constant frequency $\nu=16$\,MHz. The solid 
line is a bimodal fit to a septet of quadrupolar satellites, with positions 
constrained by Eqn.~\ref{eq:nuQm}.
The inset is a blow-up of the rightmost satellite at $B_{-3}\approx 3.36$\,T. 
}
\end{figure}

Resonances from the minority spin-down chains are marked by a negative field 
offset, comparable in absolute value to the internal field in the
majority chains.   
The signal 
could only be found in  
longitudinal fields exceeding 2~T, since for smaller external fields
its resonance frequency $\nu_0\approx
{^{59}\gamma}\big \vert \abs{B_{int}} - \abs{B_{ext}}\big\vert$ falls below 
10 MHz, outside the frequency range of our apparatus. A field sweep spectrum 
at 10~K, recorded at fixed frequency of 16~MHz, is shown in 
Fig.~\ref{fig:hsFIrev}. It consists of a multiplet of quadrupole-spit lines 
centered at $B_n$,\cite{note_on_Bm}
similar to those detected for the majority chains, with a central resonance at 
$B_0=2.792$~T. 
The two  
satellites at $B_{-2}=3.172\,\tesla$ and 
$B_{-3}\approx 3.36\,\tesla$ are however reduced in amplitude
by approximately 75\% and 90\%  
with respect to the symmetric peaks at $B_{2}$, $B_{3}$, respectively. 
The signal loss at $B_{ext}\ge B_{-2}$ is due to the approach of 
the FI-FM transition at $B_{c2}\approx 3.5$\,T,
which is accompanied by a notable relaxation phenomenon 
that partially wipes out the signal intensity over a wide 
field interval close to $B_{c2}$  (see below).
The rightmost satellite at $B_{-3}$  
is furthermore displaced from the expected position 
by approx.\ -2\,mT (figure inset), a value
well accounted for by the demagnetization field, since the 
macroscopic magnetization significantly deviates from the FI plateau at 
$H\!=\!\mu_0^{-1}B_{-3}$ (Fig.\ \ref{fig:magneto}). 
The quadrupole coupling parameter is estimated as $\nu_Q=1.97(1)$~MHz from 
the line spacing, $\nu_Q={^{59}\gamma}(1+K_c)(B_{n-1} -B_n)$, 
and is slightly, but significantly smaller 
than in the spin-up chains. 

\begin{figure}
\includegraphics[width=\columnwidth]{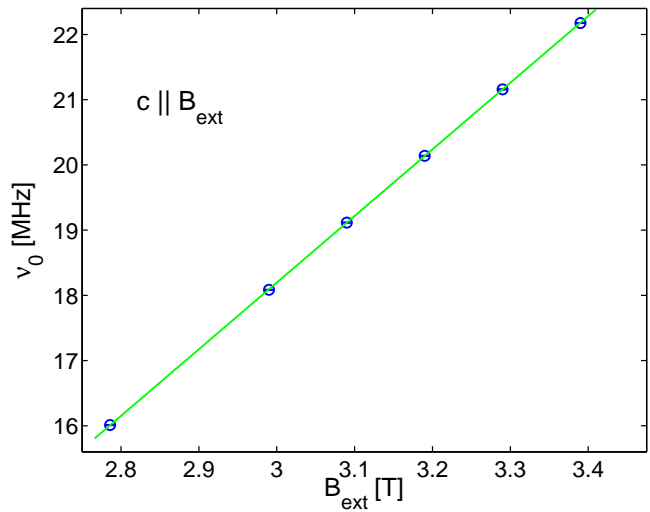}
\caption{(Color online)\label{fig:gamma_rev}
Central resonance frequency $\nu_0$ of the
spectra from the minority spin-down FI chains, as a function of  
$\B_{ext} \parallel c$. The solid line is a fit to Eqn.~\ref{eq:linear}. }
\end{figure}

The resonance frequency of the central line follows a similar linear 
dependence on $B_{ext}$ (Fig.~\ref{fig:gamma_rev}),
with slope $a_{\downarrow}=10.22(1)$~MHz/T, and a negative intercept 
$b_{\downarrow}=-12.48(3)$~MHz (Eqn.~\ref{eq:linear}). 
The former coincides within error with $a_{\uparrow}$, whence the same 
chemical shift value $K_c\approx 0.01$ is 
obtained. 
From $b_{\downarrow}$, the internal field at Co\,I in the minority chains is 
estimated as $B_{int\downarrow}=-1.221(4)$~T.

\begin{figure}
\includegraphics[width=\columnwidth]{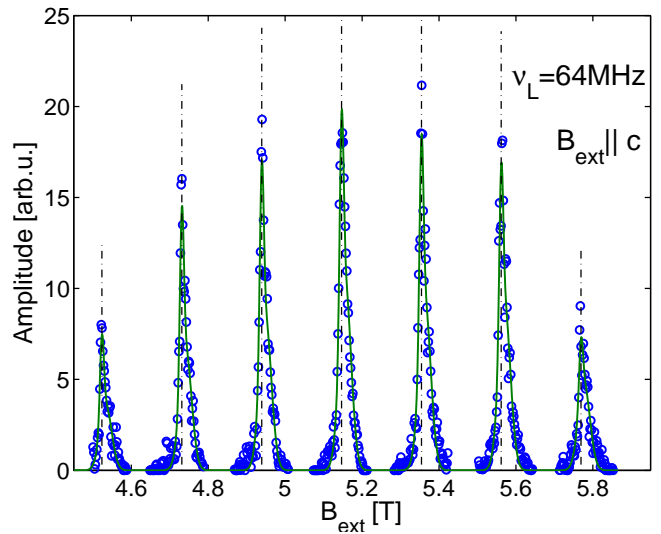}
\caption{(Color online)\label{fig:FMlines}
Field-swept ($\B_{ext}\!\parallel\! c$) \Nuc\ spectrum in the FM phase at 10~K,
recorded at a constant frequency $\nu=64$\,MHz.  
Solid lines are bimodal fits of the satellite peaks, vertical
dash-dotted lines mark their centers of gravity.}
\end{figure}

The negative-offset resonance lines 
from the minority spin-down chains completely disappear in external fields 
$B_{ext} > 3.5\,\tesla$, while only a positive-offset septet can be detected,
that is qualitatively similar to the 
one from the majority FI chains below 3.5\,T and with an identical 
quadrupole splitting $\nu_Q=2.11(1)$~MHz
(Fig.~\ref{fig:FMlines}).
The integrated amplitudes of the positive-offset spectra above 
and below 3.5\,T,  
and proportional to the number of \Nuc\ nuclei excited in spin-up chains,
have a ratio very close to 3:2.
These facts indicate that all the spin chains are now aligned parallel to 
the external field, i.e.\ the system is in a FM state.

The positive-offset septet from the spin-up chains could be 
followed at 10\,K while increasing 
$B_{ext}$ in steps of approx.\ $0.1\,\tesla$ 
across the field-induced FI-FM transition, without losing the signal. 
The resonance frequency of the central line is 
plotted vs.\ $B_{ext}$ 
in Fig.~\ref{fig:gamma}. Experimental points clearly
follow two different straight lines above and below 3.5\,T 
(Eqn.~\ref{eq:linear}), which signals the occurrence of a transition at 
a critical  field
$B_{c2}=3.50(5)\,\tesla$. 
Line parameters on the FM side are fitted as $a^{FM}=10.23(1)$~MHz/T, 
in good agreement 
with the $a$ values in the FI phase, and a reduced 
$b^{FM}=11.40(1)$~MHz, corresponding to a step-like 
decrease of the internal field, $B_{int}^{FM}=1.115(2)\,\tesla$.

The transition is also accompanied by an abrupt broadening 
by a factor of about 2 of the satellite
lines above $B_{c2}$. A closer inspection of 
the resonance lines reveals, on both sides of the 
transition, an asymmetrical bimodal shape with a sharper peak and a broader 
negatively-shifted shoulder, that is more pronounced in the FM phase. Line 
shapes and widths, in contrast, are independent of satellite order 
$m$. These facts indicate that the observed inhomogeneous broadening 
is magnetic in origin. The increase in line width $\Delta \nu_L$ above $B_{c2}$
qualitatively agrees with a dominant line broadening from 
the demagnetization field, which is proportional to its net magnetization, 
and inhomogeneous throughout the sample due to its non-ellipsoidal shape.
The agreement is not quantitative, however, since a step in $\Delta \nu_L$ 
by a factor of 3 across $B_{c2}$ is expected from the corresponding step in 
the macroscopic magnetization.
Such a discrepancy suggests the presence of an extra broadening 
mechanism, relatively more important in the FI phase.

\vspace{-1em}
\subsubsection{Transverse applied fields}
\label{sec:results.spectra.trans}
\vspace{-1em}

\begin{figure}
\includegraphics[width=\columnwidth]{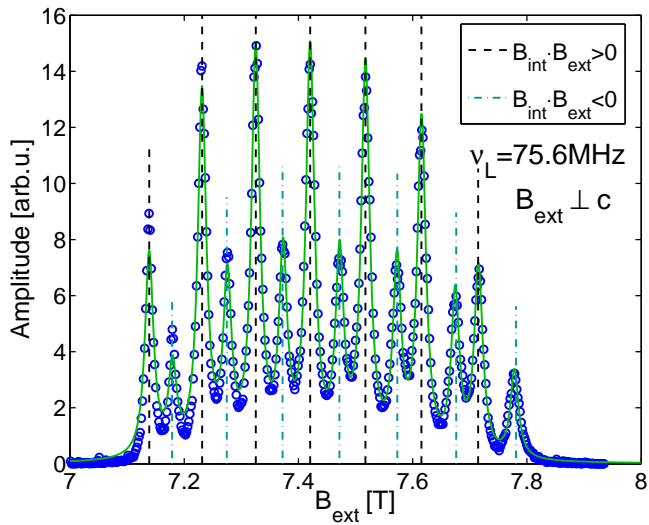}
\caption{(Color online)\label{fig:transverse}
Field-swept spectrum at 10~K,  recorded at a constant frequency of 75.6\,MHz,
 in a nominally transverse geometry ($\B_{ext}\!\perp\! c$). The solid line
is a global fit to two septets of Lorentzian lines with positions constrained 
by Eqn.~\ref{eq:nuQm}. Vertical dashed and dash-dotted lines mark peak 
positions for the majority and minority spin chain signal, respectively.}
\end{figure}

\begin{figure}
\includegraphics[width=\columnwidth]{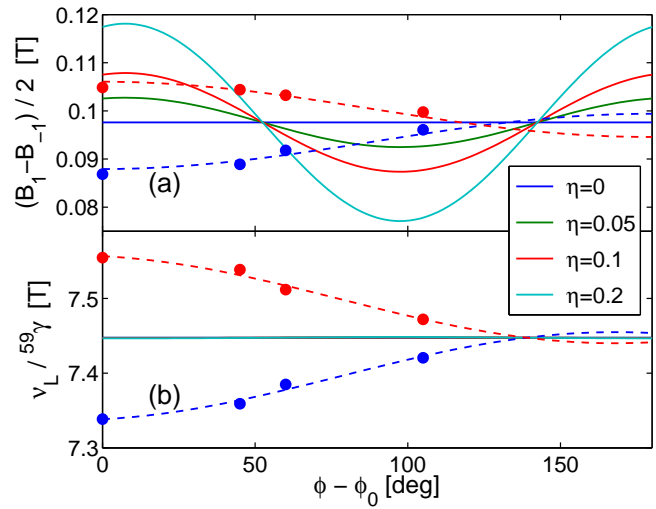}
\caption{(Color online)\label{fig:azimuth}
Symbols: experimental first order quadrupole splitting (a) and central peak 
position (b), 
in field units, as a function of the azimuthal 
angle $\phi$ in the transverse geometry. Dashed lines are guides to the eye. 
Solid lines: simulations 
with $\theta\!=\!81^\circ$, corresponding to $\B_{ext}\perp c$ 
(see text and Eqn.~\ref{eq:nuQm}), for increasing $\eta$ values.
}
\end{figure}

The crystal was also mounted
  with its $c$ axis perpendicular to the external field, in order to study the transverse components of the EFG 
and $\hat K$ tensors in an azimuthal scan of $\B_{ext}$ in the $ab$ plane. 
A typical field-sweep 
spectrum at 10~K, recorded at a fixed frequency $\nu_L=75.6$~MHz, is shown in 
Fig.~\ref{fig:transverse}. The spectrum consists of two septets of relatively 
sharp 
quadrupole-split transitions, with positive ($B_0<\nu_L/{^{59}\gamma}$) 
and negative offsets ($B_0>\nu_L/{^{59}\gamma}$) of their respective central 
lines. The integrated amplitudes of the positive- and negative-offset 
multiplets are in a ratio very close to 2. 
Following the above arguments, the two 
multiplets 
originate from the majority spin-up and minority spin-down chains, 
respectively, in the presence of a sample misalignment of the order of
a few degrees, yielding a longitudinal field
component of order a few hundreds mT, 
strong enough to induce FI magnetic order in the 
sample.\cite{note_on_powders} 
The effect of sample rotation around an axis nominally coincident with the 
crystallographic $c$ axis is summarized in Fig.\ \ref{fig:azimuth}, showing the dependence on the azimuthal angle $\phi$ of the central lines $B_{0\uparrow}$, $B_{0\downarrow}$, and of the first-order quadrupole splittings, 
calculated as 
$(B_{-1}-B_{1})/2$. \cite{note_on_Bm}
Overlaid on the experimental points, the figure also shows for comparison 
simulated shifts (Fig.\ \ref{fig:azimuth}a) 
and splittings (Fig.\ \ref{fig:azimuth}b) predicted by Eqs.~\ref{eq:nuQm} 
and \ref{eq:ABC} for $\theta=81^\circ$ (the angle 
between $c$ and $\B_{nuc}\approx\B_{loc}$, resulting from a vector addition
of $B_{ext}=7.45\,\tesla$ with a perpendicular $B_{int}\approx 1.2\,\tesla$),
and rhombicity factors $\eta$ of order 0.1.
Clearly, the experimental splittings do not follow
the predicted angular periodicity of $\pi$, rather showing a $2\pi$ period 
(Fig.\ \ref{fig:azimuth}a). 
Moreover, the quadrupolar shift is 
negligible even for large $\eta$ values (Fig.\ \ref{fig:azimuth}b), 
since it is a second order term in $\nu_Q/\nu_Z$.
Therefore, the entire azimuthal dependence of both quantities must be due to a 
slight misalignment between $c$ and the rotation axis, as well as between the 
latter and the plane perpendicular to $\B_{ext}$, indicating that 
$\eta$ is vanishingly small. 

The spectrum of Fig.~\ref{fig:transverse} is fitted to two septets of 
Lorentzian lines, with constrained positions calculated as functions of 
external and internal fields, chemical shift, quadrupole interaction, and a 
geometric angle $\alpha$ between the $c$ axis and $\B_{ext}$. In the fit, 
the in-plane chemical shift tensor component $K_a$,
$\alpha$, and $\nu_{Q\uparrow}$, $\nu_{Q\downarrow}$ 
(which are in principle independent for the two chain types) were treated as 
free variational parameters, while 
$B_{int\uparrow}$, $B_{int\downarrow}$, and $K_c$ were 
kept fixed to the
values determined by NMR with $c\parallel B_{ext}$ 
(longitudinal geometry). A further 
constant parameter, 
$\rho\equiv d\psi /dB_{ext} = 3.4(2)\!\times\!10^{-3}\,\tesla^{-1}$,
accounting for a field-dependent tilting angle $\psi$ of the electronic 
moments, was calculated from the $ab$-plane macroscopic susceptibility, 
determined by SQUID 
magnetometry as $\chi_\perp =1.8(1)\!\times\! 10^{-2}\,\mu_B/\tesla$. 
The fit yields a negative $K_a=-6.7(2)\!\times\! 10^{-3}$, and 
$\nu_{Q\uparrow}=\nu_{Q\downarrow}=2.11(1)$\,MHz,  
coincident with the quadrupole line splitting of the majority signal  
in the longitudinal geometry, and so is in disagreement
with the value determined for the minority chains in large longitudinal fields.

\vspace{-1em} 
\subsection{Nuclear relaxations}
\label{sec:results.relax}

Nuclear relaxations were systematically studied on the central quadrupole line 
($m=0$) in the longitudinal 
geometry ($\B_{ext}\parallel c$) at 10\,K as a function of the external field. 
The longitudinal nuclear polarization recovers equilibrium in good 
agreement with Eqn.~\ref{eq:sredfield}, indicating that spin-lattice 
relaxations are actually driven by magnetic fluctuations, as expected in this 
system. 

\begin{figure}
\includegraphics[width=\columnwidth]{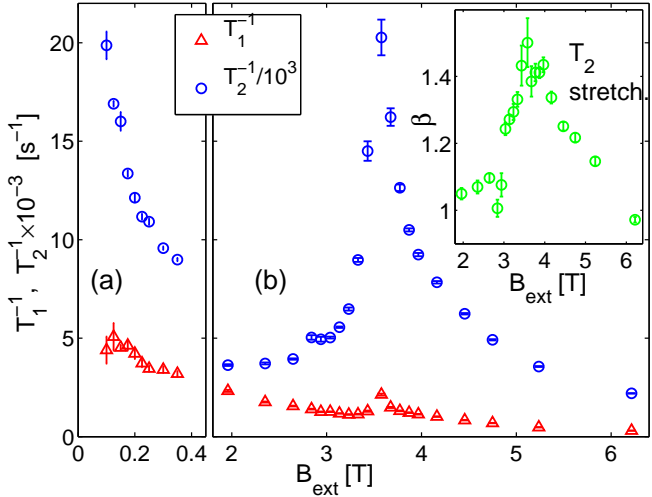}
\caption{(Color online)\label{fig:relax10K}
Spin-lattice (triangles) and spin-spin relaxations (bullets) measured on the 
central resonance of the positive-offset septet at 10\,K as a function 
of $\B_{ext}\parallel c$, (a) on approaching the FI-MPDA transition 
from above, and (b) across the FI-FM transition.
Inset of panel b: stretching exponent $\beta$ of the 
spin-spin relaxation function vs.\ $B_{ext}$. In the figure  
$T_2^{-1}$ rates are scaled by a 
multiplication factor of $10^{-3}$ for clarity. 
}
\end{figure}

Figure \ref{fig:relax10K}b 
displays spin-lattice and spin-spin  
 relaxation rates of the 
positive-offset septet at 10\,K over the 2-6~T field span, an interval 
comprising the FI-FM transition $B_{c2}$.
Spin-spin rates $T_2^{-1}$ are larger than spin-lattice rates 
$T_1^{-1} \equiv 2W$ (Eqn.~\ref{eq:sredfield})
by at least 3 orders of magnitude, and the two relaxations clearly 
follow quite different behaviors vs.\ $B_{ext}$. The former exhibits a peak at 
$B_{c2}$,
with such a large maximum value that it nearly leads to the disappearance 
(wipeout) of the signal at this temperature. In contrast, $T_1^{-1}$ 
exhibits a monotonic decrease with increasing fields, except for a 
shallow feature at $B_{c2}$.
Moreover, the time dependence of the spin-spin relaxations exhibits a marked 
non-exponential decay 
over a wide field interval across the FI-FM transition.
The signal decay is best fitted to a stretched exponential function,
$A(\tau)=A_0\exp[-(2\tau / T_2)^\beta]$, with stretching exponent $\beta$
peaked to a value of 1.5 at $B_{c2}$, i.e.\ the intrinsic lineshape 
(related to the spin-spin relaxation function by Fourier transform) is 
nearly Gaussian close to the transition.

The difference in magnitude and field dependence of the $T_1^{-1}$ and 
$T_2^{-1}$ rates is not peculiar to the longitudinal orientation,
as it can 
be qualitatively reproduced with the $c$ axis rotated by $\alpha = 45^\circ$ 
in $B_{ext} = 4$-7~T (not shown),
a field interval encompassing the FI-FM transition in that geometry.
This indicates that the  behavior of $T_1^{-1}$ and 
$T_2^{-1}$ vs.\  applied field is essentially invariant with respect to 
the orientation of the nuclear field $\B_{nuc}\approx\B_{loc}$ relative to 
the $c$ axis (note that $B_{ext}\gg B_{int}$ in this case, so that the 
$\B_{loc}$ direction is approximately that of $\B_{ext}$).

Perfectly consistent data are recorded from nuclei in the minority 
spin-down chains, for $B_{ext}$ approaching $B_{c2}$ from below. 
A qualitatively similar behavior of the two kinds 
of relaxations is also observed at 10~K  close to the other metamagnetic 
transition, i.e.\ from the FI to the 
MPDA phase on decreasing the longitudinally applied $B_{ext}$ down to  
$B_{c1}\approx 0.1\,\tesla$ (Fig.~\ref{fig:relax10K}a). 
Here again $T_2^{-1}$ is orders of magnitude larger than $T_1^{-1}$ and 
increases as $B_{ext}$ approaches the critical field, 
while $T_1^{-1}$ is non-divergent. As in the vicinity of $B_{c2}$, the 
homogeneous lineshape is quasi-Gaussian, as spin-spin relaxations are
fitted to stretched exponential decay forms, with $\beta$ in the range 
1.2\,-\,1.4.

\begin{figure}
\includegraphics[width=\columnwidth]{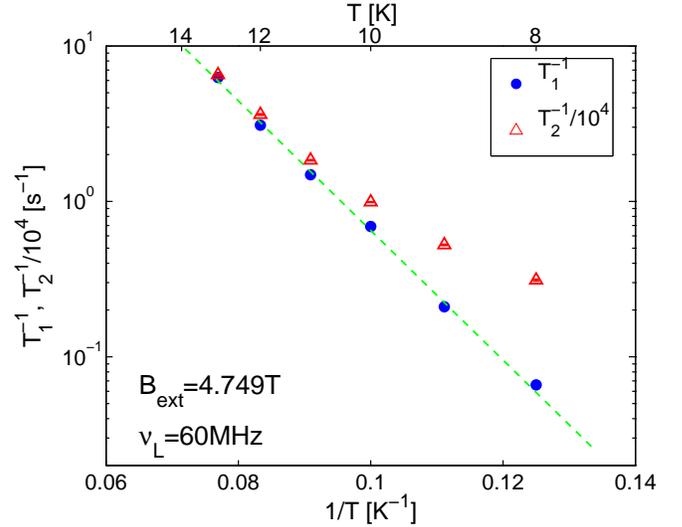}
\caption{(Color online)\label{fig:relax_vs_T}
Spin-spin (triangles) and spin-lattice relaxation rates (bullets) measured in the FM phase on the central resonance ($\nu_0=60$\,MHz, $\B_{ext}=4.749\,\tesla$ parallel to $c$) as a function of temperature. The dashed line is a fit of 
$T_1^{-1}$ to an Arrhenius law. In the figure  
$T_2^{-1}$ rates are scaled by a
multiplication factor of $10^{-4}$ for clarity. }
\end{figure}

Spin-lattice relaxations were also measured as a function of 
temperature and applied field, both in the FI and FM phase. Results of a 
typical temperature scan, performed in a longitudinal field of 4.749\,T 
(i.e.\ in the FM phase) are shown in Fig.\ \ref{fig:relax_vs_T}. 
The figure also shows the temperature dependence of $T_2^{-1}$ for comparison.
The spin-lattice rate clearly follows a thermally activated behavior  
$T_1^{-1}=T_{\infty}^{-1} \exp{-\Delta / T}$, 
 as also found by other authors. \cite{japan_nmr} 
The Arrhenius law is obeyed
 over a temperature range of 
$8-13$\,K at this particular field,
corresponding to a $T_1$ variation by more than two decades,
with activation energy estimated as $\Delta = 96(2)\,\K$.
Above this temperature interval the signal is lost due to an exceedingly short 
$T_2$, while experimental $T_1^{-1}$ points exhibit excess relaxation 
at $T< 8\,\K$, indicating that the relaxation mechanism responsible 
for the Arrhenius behavior becomes unimportant at low temperature, and it is 
shunted by other more effective relaxation channels.
Spin-spin rates seemingly saturate at low temperature
and also exhibit a steep temperature dependence at $T>8\,\K$ as well,
although the evolution is clearly not Arrhenius-like.

Thermally activated $T_1^{-1}(T)$ were detected at {\em all fields}, 
irrespective of whether the system exhibits FI or FM order.
The field dependence of the activation energy 
$\Delta(B_{ext})$, as well as its interpretation in terms of the elementary 
excitations of an Ising spin chain system, will be reported elsewhere. 

\section{Dipolar field calculation}
\label{sec:dipfield}

A calculation of the total dipolar field at the Co\,I site
is required for a quantitative analysis of our experimental internal field 
values. 
We report such a calculation here
for each of the three magnetic environments, 
majority FI-, minority FI-, and FM-ordered chains. 
The contribution of nearby moments (referred to as the proper dipolar field 
$B_{dip}$) 
 is calculated by summing up the individual dipolar fields 
 from Co\,II spins $\SS_j$ at positions $\R_j$ inside a Lorentz sphere 
with a radius of several unit cell radius centered at a Co\,I site
$\R_0\!=\!0$,  
$\B_{dip}=\sum_{j}g\mu_B (3\R_j\R_j\cdot\SS_j -r_j^2\SS_j)/r_j^5$,
while spins outside the Lorentz sphere are treated as a macroscopic 
magnetic moment density, giving rise to 
a demagnetization field $B_{dem}$ 
inside the sphere, as is customary.
The latter is strictly constant throughout the sample volume 
only 
in the case of an 
ellipsoidal sample surface, 
whereby it is simply proportional to the saturation 
magnetization $M_s$ and to the difference in demagnetizing factors of 
the inner Lorentz sphere $1/3$ and the outer sample surface $N$, 
$B_{dem}=4\pi\,(1/3-N) M_s$.  
For a rotational ellipsoid magnetized along its
principal axis $c$, the demagnetizing factor $N_c$ is 
calculated as 
\cite{landau_lifshitz}
\begin{eqnarray}
\label{eqn:demag}
N_c &=& \frac{(1  -  \epsilon^2)
( \tanh^{-1} \epsilon - \epsilon )}
{\epsilon^3}\\
& & \epsilon   \equiv  \sqrt{1 - (r_a/r_c)^2} \nonumber 
\end{eqnarray}

A non-ellipsoidal sample may be 
approximated by an effective ellipsoid
accounting for the mean $B_{dem}$,
while deviations from the ideal shape gives rise to 
a distribution in 
demagnetization fields, contributing only to the linewidth. 
We assume effective ellipsoid  radii $r_a=r_b$, 
$r_c$ equal to half of the sides of our square-base 
parallelepipedal specimen as a best ellipsoidal  approximation.
The choice of $r_a$ and $r_c$ does not critically affect the 
determination of $B_{dem}$, since $N_c$ depends  weakly on $\epsilon$ in the 
case of an elongated sample 
(e.g.\ $\partial N_c /\partial \epsilon \approx 0.3$ for 
our crystal).
 For instance, a 10\% error on $\epsilon$ would result in an 
error $\Delta B_{dem}$
of order 5 and 2 mT in the FM and FI phases, respectively.

The results of our calculations are summarized in table 
\ref{tab:dip} and compared with experimental results. Both the dipolar 
sums and the demagnetization field were calculated by assuming the
Co\,II moments saturated to $\pm g\VEV{S_z}\mu_B$ along $c$  in all the 
three types of magnetic chains, with a $\pm$ sign as appropriate for the chain 
order, and identical
absolute moment values estimated as $g\VEV{S_z}=5.25(3)$ (in units of $\mu_B$)
from an extrapolation of $M(H)$ data in high field.
Lattice parameters were taken as 
$a=9.061$\,\AA , 
$c=10.367$\,\AA\ 
as reported in the literature at 40\,K. \cite{jsolstatechem}
Numerical convergence of $B_{dip}$ could be attained with a Lorentz sphere 
radius of the order 30 of unit cells.
Calculated and experimental values 
differ by less than 2\% in all cases.
Agreement should be considered satisfactory in view of the uncertainties in the
magnetic moment (of order 1\%) and the demagnetization factor.
This confirms that the internal field is dipolar in origin, 
while transferred hyperfine contributions, if they occur, are negligible.

\begin{table}[t]
\caption{\label{tab:dip}Calculated dipolar sum $B_{dip}$, demagnetization 
field $B_{dem}$, and total dipolar field $B_{tot}\equiv B_{dip}+B_{dem}$  as a 
function of magnetic chain ordering, compared with experimental $B_{int}$ 
values.}
\begin{ruledtabular}
\begin{tabular}{ccccc}
Order  & \mbox{$B_{dip}$} (T)& \mbox{$B_{dem}$} (T) & \mbox{$B_{tot}$} (T) 
& exp.\ (T) \\ 
\hline
\mbox{FM}             & ~0.997  & 0.137  &  ~1.134 & ~1.115(2)  \\
\mbox{FI$\uparrow$}   & ~1.122  & 0.046  &  ~1.167 & ~1.178(2)  \\
\mbox{FI$\downarrow$} & -1.246 & 0.046  & -1.201  & -1.221(4)  \\

\end{tabular}
\end{ruledtabular}
\end{table}

\section{Discussion and conclusions}
\label{sec:discussion}

We organize the discussion of our experimental results into separate sections 
for sake of clarity, each one focusing on one main conclusion of the paper.

\vspace{-1em}
\subsection{Interactions of \Nuc: evidence against a direct Co\,I-Co\,II superexchange path} %

Our NMR study on a single-crystal provides a determination of the chemical 
shift and the EFG tensors at the Co\,I site.
The chemical shift is less precise due to the
uncertainty in the zero-shift \Nuc\ reference \cite{note_on_gamma}
and to
the presence of a large internal field which complicates the unraveling of 
$\hat K$ components 
(see Appendix \ref{sec:nuc}).
The absolute magnitude of $K$, of order 1\%, though sizable in absolute 
terms, is however only slightly larger than the typical values reported
for diamagnetic complexes of cobalt. \cite{Co59_ref} 
The relatively small value of $K$ agrees with the known large crystal field 
splitting at the octahedral I site, of order 0.65\,eV, \cite{burnus}
since a near-lying crystal field excited multiplet usually gives rise to a 
large chemical shift via the van-Vleck mechanism, as often encountered  
in non-magnetic transition metal ions. 

The EFG exhibits cylindrical symmetry around $c$, 
in accordance with the fact that $c$ axis is a $C_3$ symmetry element of the 
point group of Co\,I.
Consistent values of the quadrupolar frequency 
$\nu_Q$ (proportional to the EFG) are estimated 
in different chain types and experimental geometries,  
with the notable exception of the minority chains in external longitudinal 
fields approaching the FI-FM transition value $B_{c1}$,
where an EFG deviation of order -7\% is found, well
beyond the experimental uncertainty. 
We argue that a magnetoelastic coupling of the \Cot\ 
ions in counter-oriented minority chains might be the source of the observed 
EFG reduction, as already proposed for other cobalt-based compounds studied by 
 M\" ossbauer spectroscopy. \cite{magnetoelastic}

The internal field $B_{int}$ at the non-magnetic Co\,I site can be accounted 
for by purely dipolar interactions with the surrounding Co\,II spins, 
within an absolute 
accuracy of 2~mT in all of the three chain types.
This result, along with the experimental value of $\hat K$, sets a very 
stringent upper limit to a transferred 
hyperfine field at the nucleus. 

The contact hyperfine field may originate from both a tiny on-site magnetic 
moment on Co\,I, as that invoked 
in a neutron scattering refinement, \cite{earlyneutrons_FI} 
and from the coupling to nearest neighbors Co\,II spins.
These two contributions might partially cancel out, since either sign is 
possible.\cite{yoshie} 
Additional inter-chain contributions are negligible in view of the large 
chain spacing. 
Nevertheless, 
we can safely rule out an on-site term, since a residual on-site 
moment would require an admixture of the $t_{2g}$ with
the excited $e_g$ orbitals, in contrast with the large crystal field 
excitation energy implied by 
the observed moderate chemical shift value.
For instance, a huge chemical shift $K\approx 5$ was 
reported  
for the resonance of $^{141}$Pr in low-spin Pr$^{3+}$, in combination with a 
small fractional 
moment $\approx 0.01\,\mu_B$ on the  Pr ion. \cite{kay_prl} 
Therefore, only the transferred contribution is to be considered.

An intra-chain transferred hyperfine term $B_{hf}$ would offset 
the total dipolar field $B_{tot}$ (Tab.\ \ref{tab:dip}) by 
$\pm \abs{B_{hf}}$ for the FI$\uparrow$ and FM chains, and by 
$\mp \abs{B_{hf}}$ for the FI$\downarrow$ chains, 
depending on the relative sign of $B_{tot}$ and $B_{hf}$.
However,  a finite $B_{hf}$ biasing term does not improve the
agreement between the calculated $B_{tot}\pm B_{hf}$ and 
the experimentally determined internal fields. We conclude therefore that the 
slight discrepancy between the experimental and the calculated internal fields 
must be due to other sources of error, and that the contact field, if present, 
is much smaller than 2\,mT.

A negligible hyperfine field at the Co\,I site 
is an unexpected result. 
Transferred hyperfine fields at nuclei of non-magnetic ions in magnetic 
compounds of transition metals are typically of the 
order of several hundreds mT or more, unless perfect 
cancellation occurs at a symmetric site of an AF structure, 
which is not the case in \CaCO. For instance, 
the transferred hyperfine field at $^{139}$La in the FM manganite
La$_{1-x}$Ca$_{x}$MnO$_3$ equals 3\,T, i.e.\ 
approximately 10\% of the on-site hyperfine field at the Mn sites,
where the magnetism resides.\cite{allodi_ca50} 
Here, a virtually vanishing contact field at Co\,I indicates that
there is no significant hybridization between the \Cot\,(I)
and the \Cot\,(II) wave functions
either directly or through an oxygen bridge. 

The mechanisms underlying exchange and hyperfine transfer are
similar, although not exactly identical.\cite{RadoSuhl_J,RadoSuhl_hf}
Therefore, 
the lack of a detectable transferred hyperfine field 
strongly suggests that 
\Cot\,(I) ions do not participate in the exchange interaction $J_1$ 
between two neighboring on-chain \Cot\,(II) spins. 
We stress that this finding constitutes experimental evidence against the 
currently 
accepted theoretical model by 
Fr\'esard {\it et al.},
who instead proposed an exchange path across Co\,I assuming a large 
overlap of 
\Cot\,(II) with \Cot\,(I) wave functions, yielding a 
direct integral $t=1.5$\,eV.\cite{fresard} 

\vspace{-1em}
\subsection{Nature of the metamagnetic transitions: perfect Ising behavior.}

At 10\,K two types 
of field-dependent magnetic structures are 
detected, 
namely perfect FI and FM ones, 
in accordance 
with the two plateaus in magnetization data at the same temperature.
The two structures are clearly identified by distinct local fields at 
the Co\,I sites in good quantitative agreement with the dipolar 
fields calculated for the corresponding spin arrangements.
The transition from a FI to a FM phase shows up as a sharp 
step in $B_{int}$ at a threshold field 
$B_{c2}\approx 3.5\,\tesla$, revealed 
by an abrupt crossover of the resonance frequency vs.\ 
field between two distinct $\nu_0(B_{ext})$ functions (Fig.~\ref{fig:gamma}).

\begin{figure}
\includegraphics[width=\columnwidth]{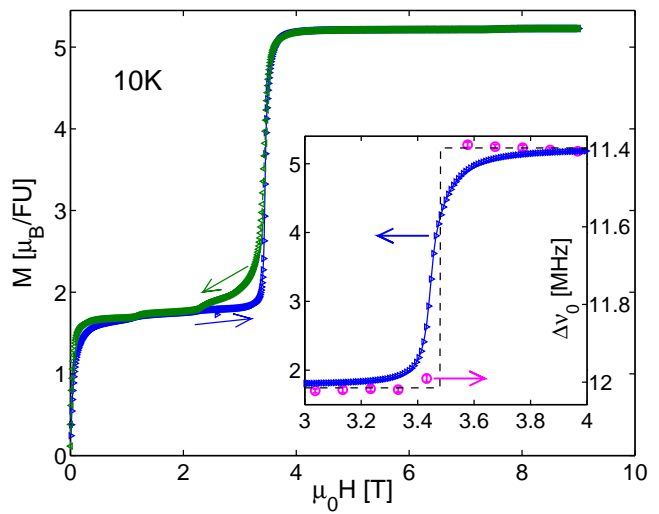}
\caption{(Color online)\label{fig:magneto}
Magnetization curves at 10~K as a function of the external field $H$ applied 
along $c$. Inset:
detail of $M(H)$ across the FI-FM transition, overlaid for comparison 
to the frequency offsets $\Delta \nu_0\equiv \nu_0 - aB_{ext}$ of the 
central NMR line of either FI$\uparrow$ or FM chains, where $a$ is the best-fit
parameter to Eqn.~\ref{eq:linear} 
(whence $\Delta\nu_0 \approx b \propto B_{int}$ within error).}
\end{figure}

Quantitative comparison between NMR and magnetization data
in longitudinal fields 
at 10\,K (Fig.\ \ref{fig:magneto}) provides an
insight into the magnetization process. 
Close inspection of the magnetization curves as a function
of the applied field $H\equiv B_{ext}/\mu_0$ 
shows that the steps appear smoother in the $M(H)$ curves than in 
the NMR frequency data
$\nu_0(B_{ext})$ (figure inset). 
Moreover, the plateaus in $M(H)$ are only approximately flat,
with a moderate field dependence that is larger in the FI state 
(for instance, $\mu_0^{-1} \partial M / \partial H \approx 0.05 \mu_B/\tesla$ 
for $1\,\tesla < \mu_0H < 3\,\tesla$).

In principle, incomplete saturation of $M(H)$ over the plateaus might be due
to either unsaturated $c$ axis components of individual moments in a 
homogeneously magnetized sample (e.g.\ due to spin canting), or 
inhomogeneous magnetization, 
although the former could not easily be
reconciled with the known strong single ion anisotropy of the system. 
NMR demonstrates experimentally that the cobalt spins are collinear
and saturated even in the vicinity of the FI-FM transition.
Proof is provided by the field proportionality coefficient $a$ 
of Eqn.~\ref{eq:linear}, 
whose experimental values for the three chain types
were found to coincide within an error $\delta a\approx 10$\,kHz/T.
The observed  value of $\partial M / \partial H$ contributes a 15~kHz/T 
slope (i.e.\ of order $\delta a$) to $\nu_0(B_{ext})$ via the 
demagnetization field, regardless of the mechanism responsible for it, and  
independent of the magnetic order in a chain. 
A hypothetical field dependent canting, however, would affect the 
dipolar field at Co\,I, which is larger than the demagnetization field
by an order of magnitude 
(Tab.\ \ref{tab:dip}). Unless we assume equal 
canting in the FM, FI spin-up, and the FI spin-down chains, which is clearly 
unphysical, the latter would produce differences in the $a$ 
coefficients of the three lines well beyond experimental errors,  
and contrary to experimental evidence. 

As seen from NMR, the system apparently 
exhibits perfect Ising behavior. 
Therefore, the rounded steps in the $M(H)$ data
at the field-induced metamagnetic transitions, 
with unsaturated magnetization over broad $H$ intervals near them, 
 must be due to a fraction of misoriented domains or spin chains.

\vspace{-1em}
\subsection{Two distinct magnetic excitations probed by $T_1^{-1}$ and 
$T_2^{-1}$} 

Further insight on the spin reversal process close to the
MPDA-FI and FI-FM transitions is provided by the 
large and divergent spin-spin relaxations, accompanied by 
much smaller and non-divergent spin-lattice relaxations.
We recall that the spin-lattice relaxation rate $T_1^{-1}$ probes 
transverse random magnetic fields fluctuating at the Larmor 
frequency $\omega_L$,
while the spin-spin rate $T_2^{-1}$ is the sum of a term proportional to 
$T_1^{-1}$ (the population term) and a secular term arising from
longitudinal random fields fluctuating at virtually zero frequency. 
\cite{abragam, slichter}
In the presence of isotropic and fast fluctuations in the so-called narrowing
limit (such that $\omega_L$ is much smaller than the roll-off frequency of the 
fluctuation spectrum) the two rates tend to coincide. \cite{note_highspin}
Here, the similar behavior of relaxations measured in 
longitudinal and oblique external fields rules out the possibility that 
anisotropic fluctuations, namely, entirely along 
the $c$ axis and hence ineffective for spin-lattice relaxations, 
 are the source of the huge difference in $T_1$ and $T_2$ time scales.
Excess spin-spin relaxation and $T_2^{-1}$ peaks 
must instead depend on fluctuations in a regime opposite to the narrowing 
limit, namely with  characteristic frequencies much lower 
than $\omega_L$ and
hence contributing only to the secular term in $T_2^{-1}$. 
Such a regime, which has already been reported in the literature for other 
oxides, \cite{allodi_wipeout}  is referred 
to  hereafter as the nearly-static limit. 

To this end, the Kubo-Tomita theory of line shape appropriate for 
continuous-wave magnetic resonance \cite{kubotomita} is modified in the 
context of pulsed NMR by 
an experimental low-frequency cutoff, 
 namely the reciprocal time window $\omega_c = (2\tau)^{-1}$ of a 
spin-echo experiment (here, $\tau$ is the time separation of the two 
excitation pulses). 
 Random fields fluctuating at frequencies  
$\omega \ll \omega_c \approx 2\times10^4\,\s^{-1}$
may be viewed as static over the excitation-detection transient 
and contribute only to the inhomogeneous line broadening, which is 
refocused by the spin echo sequence. 
Conversely, fluctuations at 
$\omega_c < \omega \ll \omega_L$ effectively relax the spin-echo 
signal.

Within the nearly-static 
limit, the  
decrease of $T_2^{-1}$ with decreasing temperature (Fig.\ \ref{fig:relax_vs_T})
then indicates 
the collapse of the fluctuation spectrum below the 
cutoff $\omega_c$ at lower 
temperatures, down to a complete freezing of the spin dynamics observed at 
$T < 5\,\K$.
It is worth noting that evidence for comparable time scales in the spin 
dynamics was also reported from $\chi'$ and $\chi''$ ac magnetic 
susceptibilities in the zero-field phase,
both showing a strong frequency 
dependence over the acoustic and subacoustic range. \cite{glassy}
The nearly Gaussian $T_2$ relaxation form is another signature of nuclear spin 
depolarization from quasi static random fields. \cite{abragam}
In our case, 
random fields are 
clearly electronic in origin,
since the calculated second
 moment of the nuclear dipolar fields yields a $T_2^{-1}$ rate smaller 
than the observed 
ones by at least one order of
 magnitude. 

The peculiar behavior of $T_2^{-1}$ rates therefore reflects the 
exceedingly slow 
spin dynamics, 
as in the case of  massive excitation modes involving the 
collective motion of several spins. 
In view of the strong Ising character of 
the system, such excitations probably consist in the coherent reversal of
 large portions of the spin chains or domains.
The slowly  
fluctuating random fields responsible for spin-spin relaxation are then 
straightforwardly 
identified with the stray dipolar fields originating from such transient 
defects in the magnetic structure.
Also 
the peaks in $T_2^{-1}$ and the stretched 
exponent $\beta$ 
at the metamagnetic transitions finds a natural explanation within this  
framework. 

A peak in $T_2^{-1}$ 
vs.\ $B_{ext}$ 
 may arise in principle from  either a field-dependent correlation 
time of fluctuations, or a peak in the random field amplitude.
In the nearly-static fluctuation regime, however, 
field dependence of the correlation time would imply {\em faster} spin 
fluctuations at the peak.
The latter 
is apparently inconsistent with the peaks 
in $\beta$ vs.\ $B_{ext}$
reaching nearly-Gaussian values of 1.5 at $B_{c1}$ and $B_{c2}$, a 
signature of a nearly frozen spin dynamics, 
following the above argument.
The $T_2^{-1}$ peak must therefore originate from a corresponding 
increase of the quasi-static magnetic 
disorder, namely a larger number of misoriented domains or chain fragments,
close to the transitions.
We note that the peaks of $\beta$ vs.\ $B_{ext}$ 
coherently 
 reflect changes in the 
spatial distribution
 of random fields on approaching the metamagnetic transitions.
 According to a 
well-known mechanism, \cite{abragam, gauss_lorentz}  
the transition from a quasi-Gaussian to a quasi-Lorentzian line shape may 
be driven by an increasing dilution of (nearly) static magnetic 
impurities or defects, and vice versa.
Thus, the larger disorder probed by $T_2^{-1}$ peaks consistently coincides
with a larger concentration of magnetic defects indicated by $\beta > 1$,
which agrees
with the shortening of the magnetic coherence length observed on approaching
 the field-induced transitions. \cite{Mazzoli09}

In contrast, spin-lattice relaxations are governed by
independent excitations, whose effect on $T_2$ is unimportant as it is 
masked by the 
relaxation channel sketched above. 
The thermally activated behavior of  $T_1^{-1}$ 
reveals a {\em gap} $\Delta$ of order 100\,K in their spectrum. A thorough 
investigation of $\Delta$ as a function of $B_{ext}$ for the different
chain types, and a fitting of the experimental energy gaps 
to a theoretical model, are the subject of a separate paper.
Here we anticipate that, according to the model, the activated spin-lattice 
relaxations are essentially driven by the spin-flip of a single \Cot\ 
moment. 

\vspace{-1em}
\subsection{Spin freezing: relevance for the 
magnetization steps} %

We conclude this discussion with a comment on the low-temperature multi-step 
behavior in $M(H)$. The magnetic structures underlying each step could not be 
accessed in our NMR study, due to the  
non-spherical shape of our sample, inducing demagnetization-dependent line 
broadening 
and offsets which prevent finer resolution of dipolar fields, as well as to
exceedingly slow spin-lattice relaxations below 5~K. 
Nevertheless, $T_2$ data at 7-15\,K reveal 
a scenario of glassy 
spin dynamics involving 
freezing of sizable 
magnetic 
aggregates and 
large quasi-static inhomogeneities close to the field-induced transitions. 
The collective arrangement of such frozen entities at lower temperature is 
unknown. We may argue however that it gives rise to complex field-dependent
micromagnetic configurations or superstructures, resulting in a multiplicity 
of local minima separated by potential barriers in the free energy, 
becoming metastable below 2~K. 
If this picture holds true, however, the physical mechanism behind 
magnetization steps should be understood in terms of thermally assisted hopping
between essentially mesoscopic metastable configurations, probably 
incommensurate with the crystal lattice, 
rather than quantum tunneling as in molecular magnets.

\section*{ACKNOWLEDGMENT}
The authors thank S.\ Carretta and P.\ Santini for helpful and stimulating
discussion. 


\appendix

\section{The \Nuc\ resonance}
\label{sec:nuc}

The resonance frequencies of \Nuc\ at the non-magnetic site I,  
for an arbitrary experimental geometry, depend on the external field and sample 
alignment in a rather complicated 
way, which deserves some comment.

The field at the nucleus $\B_{nuc}$ (proportional to the resonance 
frequency $\nu_Z=\gamma B_{nuc}$) differs from the local field 
$\B_{loc}$ at the ion site by the chemical shift, 
$\B_{nuc} = (1+ \hat K)\B_{loc}$, where the chemical shift tensor $\hat K$ is 
usually large and anisotropic in cobalt.\cite{Co59_ref} 
In a magnetic material, the local 
field is the vector composition of the external and the internal field, 
$\B_{loc} = \B_{ext} + \B_{int}$, where the internal field $\B_{int}$ 
originates from the electronic moments. The overall dependence of $\nu_Z$
is summarized by Eqn.~\ref{eq:linearB}. 

In the present case, the sources of
$\B_{int}$ are {\it i)} the dipolar coupling with neighboring Co\,II spins;
 {\it ii)} the net demagnetization field, namely, the difference between the 
demagnetization effects of the outer sample surface and a Lorentz sphere, 
which does not vanish due to the non-spherical shape of the sample; and
{\it iii)} possibly, a transferred hyperfine contribution, due to the 
polarization of wave functions.
Even in a hard easy-axis magnet like the present one, the direction of
 magnetic moments with respect to the crystal axes, hence of $\B_{int}$, 
depends on $\H \approx \mu_0^{-1}\B_{ext}$ via the finite transverse magnetic 
susceptibility 
$\chi_{ab}$ of the system, leading to a field-induced tilting of
the \Cot\ spins by a small but non-negligible angle 
$\psi \propto \chi_{ab} H_\perp$, and by 
$\approx -\psi/2$ for $\B_{int}$, if the dipolar contribution dominates.

In a crystal, \Nuc\ 
(\hbox{$I = 7/2$})
is also coupled by its quadrupole moment to the 
local electric field gradient (EFG) via a quadrupolar spin Hamiltonian of the 
form  
\begin{equation}
\label{eq:HQ}
{{\cal H}_{Q}} = 
 \frac{h\nu_Q}{6}\Big [3I_z^2 -I\left (I+1 \right) +
\frac{\eta}{2}\left (I_{+}^2 + I_{-}^2\right ) \Big ]
\end{equation}
where $I_\pm=I_x\pm iI_y$, and $I_x$ $I_y$ $I_z$ are the nuclear spin 
components in the crystal reference frame.
Here the quadrupolar frequency $\nu_Q$ is defined as a function of the nuclear 
quadrupole moment $Q$ and the main component $V_{zz}$ of the EFG tensor as 
$\nu_Q=3eV_{zz}Q/[2hI(2I-1)]$, and
$\eta=\abs{\abs{V_{yy}}-\abs{V_{xx}} }/V_{zz}$ 
is the EFG rhombicity factor. \cite{abragam} 

In the presence of a large magnetic field (either external or internal) at the 
nucleus, such that $\nu_Z \gg \nu_Q$, the quadrupolar interaction of 
Eqn.~\ref{eq:HQ} behaves as a perturbation, and splits the nuclear Zeeman 
transitions into a multiplet of $2I$ satellite lines.  
In a second-order expansion vs.\ $\nu_Q/\nu_Z$, the frequency $\nu_n$ of the $\Bra{n\!-\!1/2} \leftrightarrow \Bra{n\!+\!1/2}$ Zeeman transition
($n=-2I+1/2, -2I+3/2, \dots 2I-1/2$) depends on $\B_{nuc}$ as 

\begin{eqnarray}
\label{eq:nuQm}
\nu_n & = & \nu_Z - 
\frac{4\nu_Q^2}{\nu_Z}\left[\left(I\!+\!1/2\right)^2 -1\right] 
\left [ 2{\cal B}(\theta,\phi)-{\cal C}(\theta,\phi) \right ]  
- \nonumber \\
& & \nu_Q{\cal A}(\theta,\phi)n +
\frac{12\nu_Q^2}{\nu_Z}\left [ 4{\cal B}(\theta,\phi)-{\cal C}(\theta,\phi) 
\right ]n^2 
\end{eqnarray}

where $\theta$, $\phi$ are the polar coordinates of $\B_{nuc}$ in the crystal 
frame, and the coefficients ${\cal A}$, ${\cal B}$, ${\cal C}$ are defined as
\begin{eqnarray}
\label{eq:ABC}
{\cal A}(\theta,\phi) & = & \frac{1}{2}\left( 3\cos^2\theta -1 +
\eta\sin^2\theta \cos2\phi\right) \nonumber \\
{\cal B}(\theta,\phi) & = & \frac{1}{144}\sin^2\theta \left[ \cos^2\theta \left(9 - 6
\eta\cos2\phi + 
\eta^2\cos^22\phi\right)\nonumber \right .  \\ & + & \left . \eta^2\sin^22\phi \right ] \nonumber \\
{\cal C}(\theta,\phi) & = & \frac{1}{576}\left\{ 9\sin^4\theta 
+ 6\eta\sin^2\theta \left(1+ \cos^2\theta \right)\cos2\phi 
\right . +  \nonumber \\
 & & 
\mskip-48mu \left . 
 \eta^2\left[\cos^22\phi + 2 \cos^2\theta\left(1+ \sin^22\phi \right) + 
\cos^2\theta\cos^22\phi \right] 
\right \} \nonumber \\
& & 
\end{eqnarray}
We note that all second order terms vanish for a cylindrical EFG ($\eta=0$) 
and EFG principal axis collinear to $B_{nuc}$ ($\theta=0$).  

\section{The spin-lattice relaxation function}
\label{sec:redfield}

The approach to thermodynamic equilibrium of an ensemble of nuclear spins $I$
is governed by a closed set of rate equations for the $m$-th level population 
$N_m$ ($m=I,\dots, -I$). 
In the presence of a magnetic relaxation channel, it has  
the form \cite{narath}
\begin{eqnarray}
\label{eq:rateq}
\dot{N}_m &=& W [(I+m+1)(I-m)(N_{m+1}-N_{m+1}^{(0)})  \nonumber \\ 
&+& (I-m+1)(I+m)(N_{m-1}-N_{m-1}^{(0)}) \nonumber \\
&-& 2(I^2-m^2+I)(N_{m}-N_{m}^{(0)}) ] 
\end{eqnarray}
where $N_{m}^{(0)}$ is the population at equilibrium, 
and $W$ is the transition probability 
$\vert m\rangle \leftrightarrow \vert m-1\rangle$
between two adjacent  Zeeman levels.
In the case of quadrupole-split NMR lines with the selective excitation 
of a single nuclear transition of order $n$ ($n=I-1/2, \dots, -I+1/2$), 
the signal intensity is proportional to the population difference 
$A_n\equiv N_{n+1/2}-N_{n+1/2}$ of the two levels 
involved, rather than the expectation value 
of the total spin $\VEV{I_z}$. Solution of Eqn.~\ref{eq:rateq} 
yields a 
recovery of $A_n$ according to the superposition of 
$2I$ exponential components for generic $n$ (reduced to $I+1/2$ for 
$n=0$) with relative weights depending on the 
saturation method as well as on $n$. 

In our experiments, 
the spin ensemble was prepared 
with a fast saturation, i.e.\ with a pulse train 
much shorter than the spin-lattice relaxation time $T_1\equiv (2W)^{-1}$, 
so that populations $N_m$ of nuclear levels $m\neq n\pm 1/2$ not involved in 
the transition were practically unaffected. \cite{narath,rega}
For the central line $n=0$ of a $I\!=\!7/2$ 
nuclear species as \Nuc, the recovery law  
following such an initial condition 
is then calculated as follows.  
\begin{eqnarray}
\label{eq:sredfield}
A_0(t)= A_0^{(0)} \left ( 1 - \frac{1}{84}e^{-2Wt} \right . 
&-& \frac{3}{34}e^{-12Wt} - \nonumber \\
\frac{150}{728}e^{-30Wt} 
 &-& \left . \frac{1225}{1716}e^{-56Wt} \right ) 
\end{eqnarray}



\begin{thebibliography}{}
\bibitem{note_on_Bm} Hereafter we denote as $B_n$ the position of the $n$-th 
quadrupole satellite in a field sweep spectrum at a fixed frequency $\nu_L$, 
implicitly defined by the equation $\nu_n(B_n) - \nu_L = 0$, where 
$\nu_n(B_{ext})$ is the resonance frequency of the $n$-th satellite in an 
external field $B_{ext}$.
\bibitem{note_on_gamma}
The \Nuc\ gyromagnetic ratio is determined as
$^{59}\gamma/2\pi=10.10$\,MHz/T from  a K$_3$Co(CN)$_6$ aqueous 
solution reference, 
see e.g.\ Ref.\ \protect{\onlinecite{ref_on_gamma}}.
 In the literature there is however large dispersion in the 
reported values for $^{59}\gamma$, due to the large chemical shifts of \Nuc\ 
and  the uncertainty on a zero-shift reference. \protect{\cite{Co59_ref}}
\end{thebibliography}
\end{document}